\documentclass[conference]{IEEEtran}
\IEEEoverridecommandlockouts
\usepackage{cite}
\usepackage{amsmath,amssymb,amsfonts}
\usepackage{algorithmic}
\usepackage{graphicx}
\usepackage{textcomp}
\usepackage{listings}
\usepackage{wrapfig}
\usepackage{tikz}
\usepackage{pgfplots}
\usepackage{pgf-pie}
\usepackage{soul}
\usepackage{url}
\usepackage{hyperref}

\usepackage{xargs}   

\pgfplotsset{compat=1.18}

\definecolor{codegreen}{rgb}{0,0.6,0}
\definecolor{codegray}{rgb}{0.5,0.5,0.5}
\definecolor{codepurple}{rgb}{0.58,0,0.82}
\definecolor{backcolour}{rgb}{0.95,0.95,0.92}

\def\BibTeX{{\rm B\kern-.05em{\sc i\kern-.025em b}\kern-.08em
    T\kern-.1667em\lower.7ex\hbox{E}\kern-.125emX}}
\begin{document}

\title{Towards Infrastructure For Change Impact Analysis in Microservices: Early Findings and Prototyping \\
}

    \title{Incremental Software Architecture Reconstruction to Foster Change Impact Analysis in Microservices
}
\title{Towards Change Impact Analysis in Microservices-based System Evolution}

\author{\IEEEauthorblockN{1\textsuperscript{st} Tomas Cerny}
\IEEEauthorblockA{\textit{SIE, University of Arizona} \\
Tucson, Arizona, USA \\
tcerny@arizona.edu}
\and
\IEEEauthorblockN{2\textsuperscript{nd} Gabriel Goulis}
\IEEEauthorblockA{\textit{SIE, University of Arizona} \\
Tucson, Arizona, USA \\
ggoulis@arizona.edu}
\and
\IEEEauthorblockN{3\textsuperscript{rd} Amr S. Abdelfattah}
\IEEEauthorblockA{\textit{SIE, University of Arizona} \\
Tucson, Arizona, USA \\
amrelsayed@arizona.edu}
}

\maketitle

\begin{abstract}

Cloud-native systems are the mainstream for enterprise solutions, given their scalability, resilience, and other benefits. While the benefits of cloud-native systems fueled by microservices are known, less guidance exists on their evolution.
One could assume that since microservices encapsulate their code, code changes remain encapsulated as well; however, the community is becoming more aware of the possible consequences of code change propagation across microservices. 
Moreover, an active mitigation instrument for negative consequences of change propagation across microservices (i.e., ripple effect) is yet missing, but the microservice community would greatly benefit from it. This paper introduces what it could look like to have an infrastructure to assist with change impact analysis across the entire microservice system and intends to facilitate advancements in laying out the foundations and building guidelines on microservice system evolution. It shares a new direction for incremental software architecture reconstruction that could serve as the infrastructure concept and demonstrates early results from prototyping to illustrate the potential impact.  

\end{abstract}


\section{Introduction}
When a software system changes, we assume it improves and adapts, but it can also lead to unexpected consequences, disrupting its functionality or qualities. We typically want to mitigate such consequences early on in the development life-cycle, but dealing with a complex system might be difficult. While we often resort to testing as the key to quality assurance, testing might not detect the consequences of change. Moreover, testing of complex systems might lack complete coverage of the system or miss details in the introduced changes. 

When considering the granularity of a code change introduced to version control by the developer's commit or pull request, a code review process seems appropriate to mitigate potential issues missed when implementing the change. Tool support exists that can provide instant feedback to developers on degraded code quality, like SonarCloud \cite{SonarClo89:online} or PMD \cite{PMD33:online}. 
However, the situation fundamentally changes when we start to cope with decentralized systems. For instance, microservices decompose the system into different units of operation with specific business purposes, aiming to provide reasonable independence to their development teams to make their own decisions. It is common practice for cloud-native systems to use microservices managed by distinct teams \cite{amoroso2023one} while encapsulating knowledge in distinct microservice code repositories.  

While microservice architecture might seem like the ideal direction for enterprises, we learn from practitioners' experience that the situation is not as straightforward when evolving microservices introduce conflicts \cite{lercher2024microservice,bogner2021,bavskarada2020architecting}. Change in one microservice can break others and lead to a ripple effect.  
This is because the microservices interact and often partially overlap in the operation domain, with requirements not always perfectly divided among microservices. Unavoidably, dependencies exist across microservices \cite{closer24}.

The established practice on code review suites single microservice, and current tools, unfortunately, fit only to the single-app perspective. The community needs a little revolution in the field of microservices and should seek novel approaches to cope with change impact analysis, taking into account the entire ecosystem.   

This paper suggests abstracting the complexity of a decentralized system and its interconnection when analyzing a code change impact. 
Traditional static analysis tailored to code constructs used in microservice development frameworks alongside targeted detection and management of microservice dependencies can provide a promising system intermediate representation serving as an infrastructure for this effort. 
In particular, we propose incremental software architecture reconstruction to maintain an intermediate representation of the overall system and apply change impact~analysis~on~top~of~it.

To demonstrate it is feasible to maintain the intermediate representation throughout system evolution and use it for change impact analysis, we consider a case study on third-party microservice benchmarks, using their version control history. We implement a prototype tool for incremental software architecture reconstruction and showcase the detection of potential breaking changes and change conflicts that span across microservices. We report these conflicts with the granularity of interconnected components. In particular, 
this demonstrates how our approach could aid the review process of individual changes and help mitigate breaking changes throughout system evolution. Moreover, we illustrate the benchmark evolution trends with respect to identified breaking changes.

Incremental software architecture reconstruction as an instrument to derive system abstraction across evolution has great potential for a broad impact on the microservice community, enabling the development of new kinds of tools to assist practitioners with change impact analysis, observation of change propagation pathways, and quality assurance, considering the scope of the whole system. 

The organization of this work is as follows. Section II dives into existing work on microservice system quality assessment and evolution management. Section III describes the incremental software architecture reconstruction process used for change impact analysis, as described in Section IV. Section V constructs a proof of concept tool to assess the evolution of the microservice benchmarks, detailing the findings. Section VI concludes with a discussion of the significance of the findings.

\section{Background and Related Work} 


This paper 
positions itself in the microservices field and targets the maintainability and evolvability of cloud-native systems. In this perspective, we cannot omit established research on architecture degradation and technical debt that aims to identify anti-patterns, smells, coupling metrics, etc. However, the intended technological advancement here is change impact analysis, which is another discipline elaborated in monolithic systems. Moreover, to provide timely feedback to developers on their changes before changes get deployed, the reader must also get briefly familiar with the software architecture reconstruction process, explicitly using static analysis. The motivation of this section is pragmatic - convince the reader of an existing gap in the field and research. 

\subsection{Challenges with Microservice Evolution}

Bogner et al. \cite{bogner2021} lays great motivation on challenges faced by practitioners working with microservices. They interview practitioners to learn about challenges they perceive for their systems' evolvability. They also provide a comprehensive overview of these challenges and note that many of the maintainability problems are only partially or insufficiently addressed. For instance, API changes in microservices remain an important issue without assurance for practitioners. 

Recall what we mentioned earlier: microservices are managed by distinct teams \cite{amoroso2023one}, which means that one team does not know what the other team is doing. Next, consider that a microservice system can have hundreds of microservices interconnected. Change propagation could become a nightmare. 

We can see the great elaboration on this topic importance, specifically API change propagation, by Lercher et al. \cite{lercher2024microservice}. They interviewed practitioners and found significant overhead on communication across teams when aiming to evolve microservice API, preventing breaking the system upon changes. Their findings suggest that the leading cause of breaking changes is the weaknesses of manual error detection. Baskarada et al. \cite{bavskarada2020architecting} also learned from practitioners that organizational challenges with API changes require intensive communication and coordination. At this point, we should step in and defend developers and code change reviewers. In the first place, how could they mitigate errors related to change propagation across microservices when all they know is the microservice codebase? One should assume that the developer is not a system architect, and while some explicit dependencies between microservices can be noticed in the code (like remote calls), many dependencies may remain implicit \cite{closer24} and can be missed in the review process.

Despite the central role of API in microservices, serving as the interface and acting as a communication hub, it is essential to look beyond API evolution alone \cite{tran2021towards}. One should consider the changes driven by APIs across different layers of the architecture, which highlights the importance of accounting for great, multifaceted changes within the architecture, including mapping actual code changes to interconnected pathways.

\subsection{Mitigating Architectural Degradation: General Practice}

Informed about the issues we are facing in this field, what does the state-of-the-art have to address this? Researchers are fascinated by identifying smells and metrics to indicate that the system architecture degrades \cite{Baabad2020327} or by detecting technical debt \cite{10.1145/3512753.3512755}. Architectural degradation can be the result of uncontrolled changes that introduce potential negative effects on maintainability (i.e., anti-patterns); they accumulate and lead to a gradual erosion of the system's architecture. The reason there is such an interest in this approach is because of the tooling support on the market. SonarCloud \cite{SonarClo89:online} is an online tool free to open-source projects used by companies like eBay, IBM, NASA, and many others to improve the maintenance of their code quality. This tool has over 600 citations on Google Scholar, and its standalone predecessor, SonarQube, has almost 6 thousand citations. Researchers consider this tool as a golden standard for the problem. Yet, it only works on a single code repository and lacks support for the microservice decentralized perspective. This tool operates by finding smells in the code and associating them with the assumed cost to fix that aggregates over time to a technical debt. 


\subsection{Mitigating Architectural Degradation on Microservices}

Architectural degradation mitigation has been approached for microservices \cite{10.1145/3512753.3512755}. For instance, Tight et al. \cite{TIGHILT2023111755} aimed at detecting microservice smells across multiple system samples, and similarly, Genfer and Zdun \cite{10.1007/978-3-031-70797-1_10} mined repositories to study the architectural evolution of APIs from repository changes. While one focused on smells, the other looked at high-level architectural metrics considering API. Both seem to be straightforward applications of mitigation strategies for architectural degradation. 
The approach adopted by these works applied regular expression matching on top of the source code. While this solution demonstrates results from a shallow engineering perspective, it lacks robustness and broader potential for future extension by the community. 

When we look further for known mitigation strategies to architecture degradation \cite{Baabad2020327}, besides smells and metrics, one could consider the detection of architectural rule violations or expert assessment. System architects might define rules that must be preserved throughout the system's evolution, and these are checked when the system changes. While distinct, it is similar to smells but seems more open to the human expert involved in the change review process.  

\subsection{Human Expert and Software Architecture Reconstruction}

To assess the system by an expert, one needs to understand the system, and this is an avenue for software architecture reconstruction (SAR) \cite{walker2021automatic}. SAR assumes that constructing a simplified overview of the system can help experts understand its full scope. SAR is core to architecture verification, conformance checking, and trade-off analysis \cite{o2002software}. When we consider the simplified overview in the context of architectural descriptions (ISO 42010) \cite{ISO42010}, there exist different meanings for architectural descriptions to different experts. Each expert might come with a different viewpoint on the system. Each viewpoint may govern one or more architectural views comprising a portion of an architecture description. 

In this context, source code and other resources like requirements documents, policies, issue tracking systems, and documentation could be inputs to the SAR. If we limit the SAR to the perspective of source code, there has been success in the extraction of certain architectural views from microservices. Cerny et al. \cite{cerny24cluster} demonstrate a static code analysis-based process that uncovered reliable representation of a service view via service dependency graphs (SDG) and the context map representing data view illustrating data entities and their overlaps in microservice-specific data models. 

The SAR process, based on static code analysis \cite{cerny24cluster}, operates with an intermediate representation formed in two steps: (1) the component interconnection (component call graphs) is parsed from the source code of distinct microservices; next, (2) the distinct component call graphs are cross-connected by identified remote calls to other microservice endpoints and by data overlaps. The resulting intermediate representation can be used for various problems: to derive visual architectural viewpoints, detect smells, or even to advise large language models on how to describe specific microservice systems \cite{quevedo2024evaluating}. 

It is essential to put the engineered solutions by Tight et al. \cite{TIGHILT2023111755}, Genfer and Zdun \cite{10.1007/978-3-031-70797-1_10} into contrast with the Cerny et al. \cite{cerny24cluster}. The first two apply regular expressions on source code to look for patterns in the code to address a simple problem. However, the SAR approach uses conventional static code analysis using code parsers tailored to microservice development frameworks. Their perspective is that microservices are developed based on components like controllers, services, repositories, or data entities, and these concepts are language agnostic. The parsing approach builds an abstract syntax tree (AST) and then identifies components in the source and their interconnections; consequently, they interconnect microservices based on approximation connecting remote calls and endpoint signatures or using data overlaps. The result is an intermediate representation of the system formed from interconnected components. Given the formation of AST, any code information could be arbitrarily added to the intermediate representation; moreover, with access to the codebase, additional artifacts are added, like deployment descriptors or build files. 

Hutcheson et al. \cite{hutcheson2024software} applied SAR to bytecode (specifically GraalVM Native Image) and demonstrated in experiments a very high precision on interconnecting microservices. Moreover, Schiewe et al. \cite{schiewe2022advancing} demonstrate an augmented component parsing from the source code in a language-agnostic way, which opens up avenues for polyglot systems. 

To engage experts, the system's intermediate representation produced from the SAR process can be transformed into various visual perspectives, facilitating observability of the interconnection of microservices \cite{cerny24cluster}. For instance, Huizinga et al. \cite{10.1007/978-3-031-61816-1_15}, or Adams et al. \cite{10.1007/978-3-031-66326-0_19} illustrate that they can detect and visualize microservice smells on top of the intermediate representation resulting from SAR using static code analysis.

\subsection{Change Impact Analysis}

Next, let's consider an analogy with code-based change impact analysis (CIA). While there is marginal work on CIA for microservices where static analysis has been applied \cite{JSS-2024}, it has been successfully used in monolithic systems. Angerer et al. \cite{cia2019} describe CIA as the process of determining the potential effects of a proposed modification in a software system. It can automatically determine and systematically review the possibly impacted source code for changes. CIA is commonly performed by abstracting the system into a system dependence graph (SDG) and then following the SDG edges. SDG consists of nodes representing concrete and abstract program elements, edges encoding control, and data dependencies. An SDG is a directed graph representing different kinds of dependencies between program elements. It usually represents control-flow and data-flow dependencies, but other types, such as definition-use dependencies, are also possible.

In the prior text, we intentionally introduced an overlapping abbreviation to highlight readers' attention. In microservices, we used service dependency graphs in their service view, but in monolithic systems, the CIA discipline uses system dependency graphs. We want to emphasize the analogy that can be used to apply CIA over the decentralized microservice perspective. However, there is an important point to make. While we mentioned control and data dependencies in the microservice system intermediate representation context, there might be other dependencies to manifest. Cerny et al. \cite{closer24} identified additional microservice dependencies that could be utilized in the intermediate representation to strengthen it. However, we will not focus on this extension and leave it to future work. With the terminology of the CIA domain, we can see the component call graph augmented with two types of microservice dependencies: control and data dependencies. 

\section{Incremental Software Architecture Reconstruction (SAR)}


Although SAR can be used for various goals, as we learned in Section II.D, it typically establishes a system intermediate representation that it uses to derive answers to the varying goals. However, the shortcoming of SAR processes is that there is no awareness of the evolving aspects of the software systems. As systems evolve, the result of SAR changes~as~well.

It is instrumental for the advancement of the microservice field to establish an incremental SAR process so that system changes can be reflected, tracked, and analyzed. For cloud-native systems that are based on microservices, such a process should be aware of system decentralization. The objective is to introduce an incremental SAR (ISAR) process that serves as a framework to facilitate analysis throughout the evolution of microservice-based systems. Section IV will elaborate on this and illustrate change impact analysis on top of this framework.



\begin{figure}[t!]
    \centering
    \includegraphics[width=.4\textwidth]{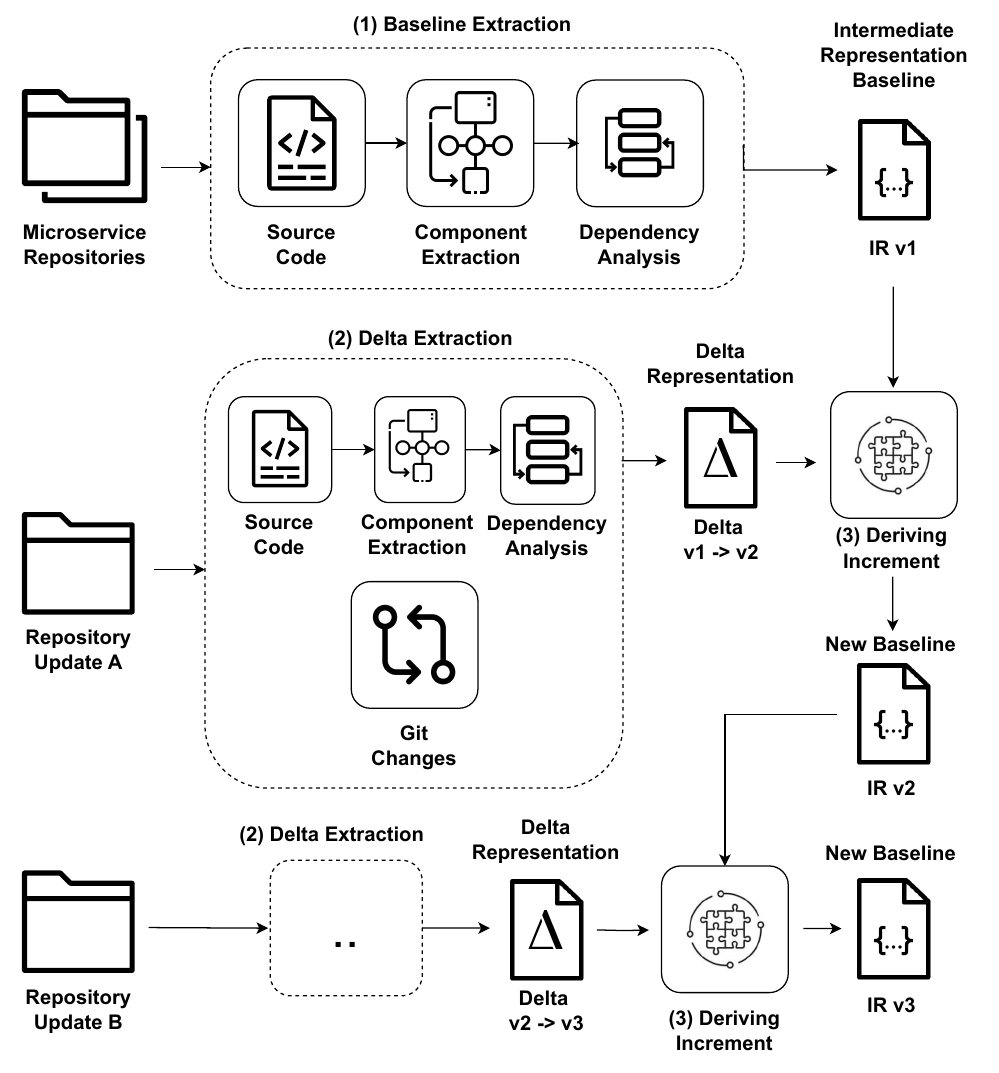}
   \vspace{-1em}
    \caption{Approach to change impact analysis using intermediate representation (IR) and deltas to maintain an up-to-date system overview}
\vspace{-1em}
    \label{fig:ToolFlow}
\end{figure}

 The proposed ISAR process consists of three stages, as outlined in Fig.~\ref{fig:ToolFlow}. It starts with the extraction of the system intermediate representation baseline (1) from microservice repositories. It is assumed that the organization has control over the decentralized system code repositories. Once the system's intermediate representation baseline is established, individual repository changes can be extracted (2) and mapped to a delta representation, encapsulating the change impact on the intermediate representation. The intermediate representation baseline is then combined with the delta representation to derive an increment (3), which can be subsequently used as a new baseline for the iterative process. The three-stage process is set to iteratively maintain an up-to-date intermediate representation baseline as the system evolves. The result of the incremental SAR process is an ordered list of intermediate representation versions with an interweaving change delta representation that encapsulates the intermediate representation changes between each version. Each transition from a baseline through a delta to the increment can be further analyzed, as elaborated in the next section on change impact analysis.
 
\subsection{Constructing System Intermediate Representation Baseline}

The system intermediate representation can act as an abstraction to oversee the system architecture from a holistic perspective. This has been previously applied to construct architectural views to assess architectural properties and foster observability and documentation by Cerny et al.~\cite{cerny24cluster}. Code analysis can be applied to the organization's decentralized repositories of the system (i.e., repository per microservice \cite{amoroso2023one}) to form the intermediate representation based on two dimensions: the dimension of a single microservice internal detail, which recognizes its components and their interconnection through method calls, and the dimension of the microservice ecosystem resulting from inter-connected component call graphs utilizing microservice dependencies.

    \begin{figure}[t!]
    \centering
    \includegraphics[width=23.8em]{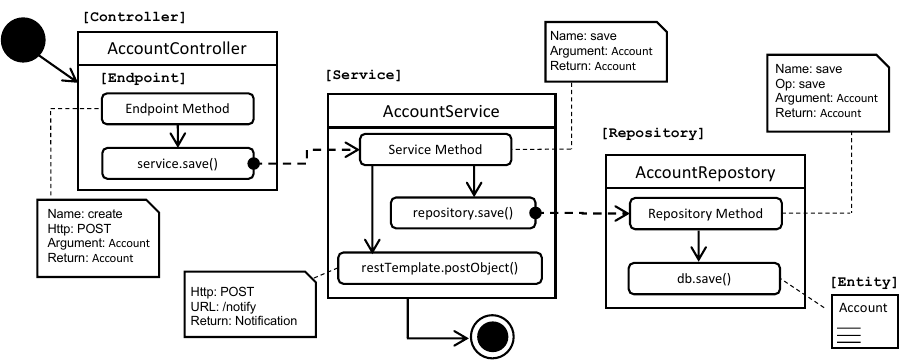}
   
    \caption{Example of the Component Call Graph \cite{cerny24cluster} illustrating how a \texttt{controller} \texttt{endpoint} connects to a \texttt{service} and \texttt{repository} that manages and initiates remote call while using 'Account' \texttt{data entity} }
    \label{fig:ccg}
        \vspace{-1em}
    \end{figure}

    \begin{figure}[t!]
    \centering
    \includegraphics[width=23.8em]{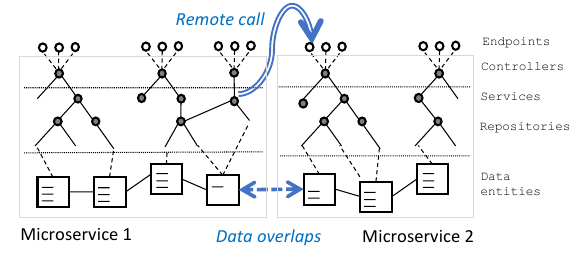}
       \vspace{-0.7em}
    \caption{Example of combining Component Call Graphs \cite{cerny24cluster} illustrating connections via dependencies given by remote calls and data overlaps. Fig. \ref{fig:ccg} is a detailed component view that is reduced to nodes in this figure}
    \label{fig:merge}
          \vspace{-1em}
    \end{figure}

The first derived dimension considers components of the three-layered architecture from which it combines component call graphs, starting with an endpoint owned by a controller component that likely utilizes a service and repository components and a particular data entity. A sample of such a component call graph is illustrated in Fig.~\ref{fig:ccg}.  A microservice would have multiple component call graphs. Across the pathway from endpoint to a data entity, all component properties send settings can be parsed and included, such as access restrictions on endpoint methods, remote calls or events generated in service component methods, etc. The internal representation of the component call graphs is in JSON format (see Zenodo$^{2}$).

Fig.~\ref{fig:ccg} illustrates three ingredients used for the second dimension. It shows the service method of account service, making a remote call, it also recognizes the endpoint signature, and finally, it uses the account entity across the component call graphs. Matching between the system endpoint and remote calls can form interconnections. Similarly, data entity similarity analysis reveals overlapping data entities across microservices. Consequently, this forms the second dimension of the intermediate representation illustrated in Fig.~\ref{fig:merge}, which is displayed with components reduced to nodes for simplification and sketched in the three-layered architecture of middleware microservices for the context. The inter-microservice connections are established through dependencies given by remote calls and data overlaps. Combining all microservice component call graphs forms the ecosystem overview. The internal format of the system's intermediate representation is a single large JSON file for the entire system.

\begin{figure}[t]
    \centering
    \includegraphics[width=22em]{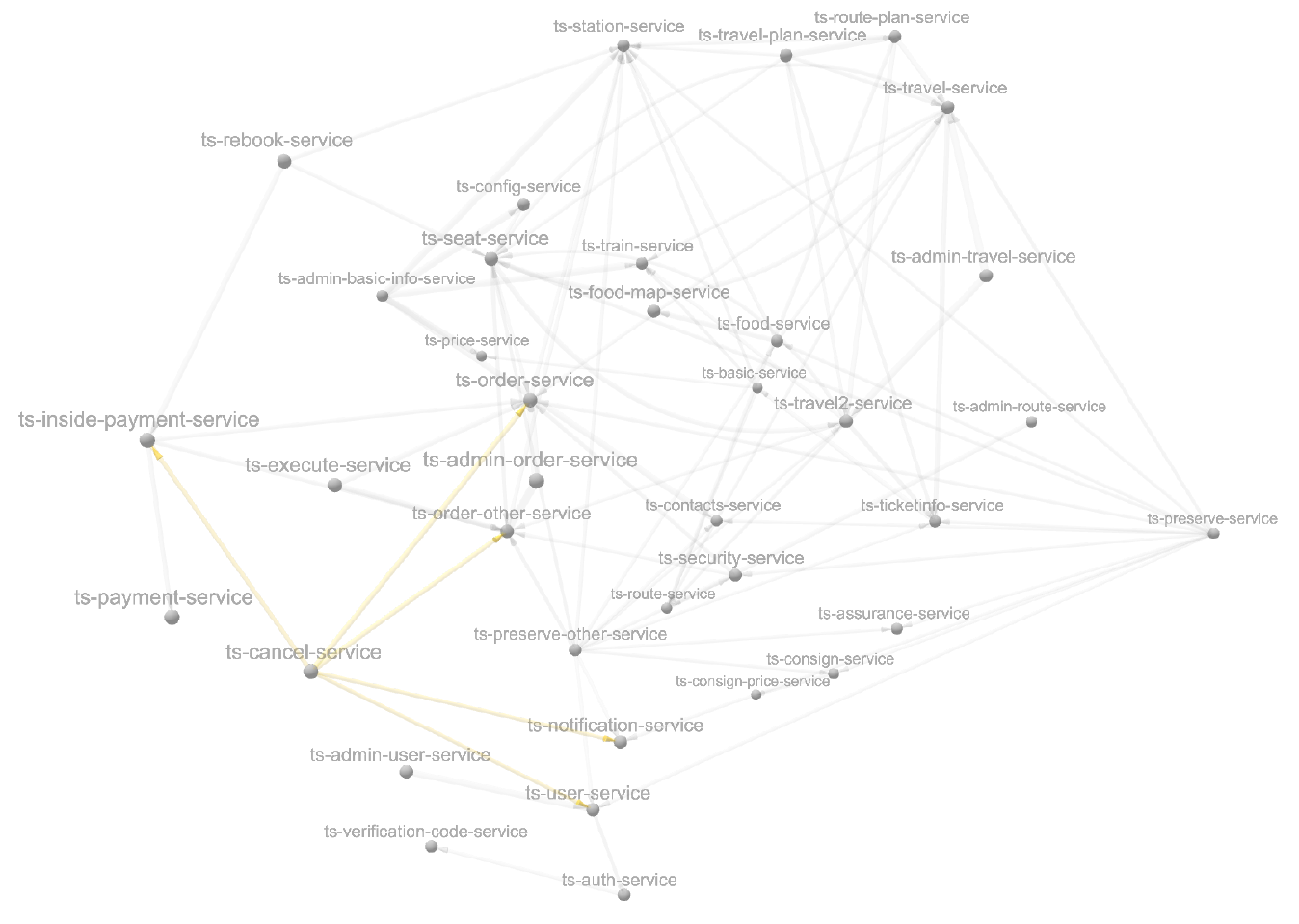}
            \vspace{-1.6em}
    \caption{Example service dependency graph of connected microservices in train\,ticket system\,\cite{Zhou2018}. Extract from the intermediate representation where a node illustrates a microservice with collapsed Component Call Graphs }
    \label{fig:system}
    \vspace{-1.2em}
    \end{figure}

The intermediate representation should be seen from a broader perspective, and the volume of information it captures. A realistic system could have hundreds of microservices. If we simplify the complexity and consider a microservice to be a node, we could manage a system like the one illustrated in Fig.~\ref{fig:system}, the train-ticket benchmark \cite{Zhou2018}. With just 47 microservices, it is complex even with the simplification; however, the intermediate representation holds the complete lower-level component granularity detail of each microservice.

The limitation of SAR processes is that they correspond to a single system version. The notion of a change initiated at a single microservice code repository must be considered. 


\subsection{Concept of a Change - The Delta}

A change in a particular microservice code repository is observed in the source control as a commit or a pull request managed by a single team. 
We refer to the difference between two versions of a single repository as a delta. A delta could contain a commit that aggregates multiple changes to single or various resources, considering their prior and new version, or it could be an aggregation of multiple commits. A delta could be a composition of multiple deltas. A delta captures a transition between two system versions. 

The structure of a delta 
contains extracted components from the source code with a label for each component indicating the type of change \texttt{[ADD/MODIFY/DELETE]} that took place. Additional metadata, such as the old and new version identifiers, are present. This makes the delta file a systematic mapping that can be overlayed onto the system's intermediate representation baseline to generate the next system version increment.



\subsection{Intermediate Representation Increment}

With the intermediate representation system baseline and a delta, we are able to reconstruct the intermediate representation version which we refer to as an increment. 
Considering a process of parsing deltas from changes rather than comparing two complete snapshots of the intermediate representation versions has performance benefits, allowing iterative increments. Moreover, if there were multiple changes across repositories of different microservices, that could factor out moving system baseline and consider validation towards the currently deployed microservices rather than the development versions.



The increment of the system intermediate representation version via delta is performed systematically utilizing the change information. There are three possibilities when considering a delta component: Components that are marked
\begin{itemize}
    \item \texttt{ADD} are new components; they will be added to the intermediate representation. 
    \item \texttt{MODIFY} are changed components.
    \item \texttt{REMOVE} are existing components that will be removed from the intermediate representation.
    
\end{itemize}

This merge process is illustrated in Fig. \ref{fig:ToolFlow}, deriving an increment, resulting in a new system baseline of the intermediate representation. It is iterative, with individual changes coming from any microservice repository.

\begin{figure}[t!]
    \centering
    \includegraphics[width=23.8em]{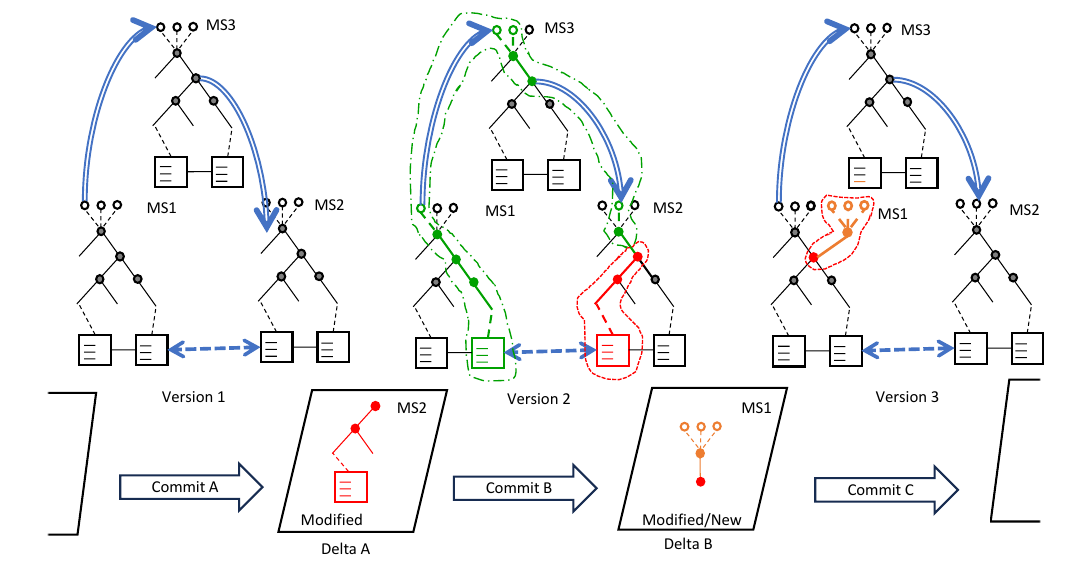}
       \vspace{-0.7em}
    \caption{Illustration of a list of evolving system intermediate representation versions with change deltas for increments. Red illustrates an update, green  potential indirect change impact; an addition is orange}
    \label{fig:isar}
          \vspace{-1em}
    \end{figure}

Fig. \ref{fig:isar} illustrates a result of the ISAR process as a list of evolving system intermediate representation versions with change deltas that increment the intermediate representation to the next version. With the incremental approach, individual changes might be factored out, and the actual change scope is well encapsulated and suits on-demand assessments if integrated into development pipelines. 

\section{On Change Impact Analysis in Microservices}




The ISAR process can serve as an infrastructure to facilitate advancements in microservice maintenance and evolution. The change impact analysis can be elaborated using the intermediate representation versions and deltas. A delta forms the increment and can be used to scope the direct and indirect 
impact of change on components in the intermediate representation. Moreover, the infrastructure allows us to introduce rules to better target specific changes with greater potential for conflicts or breaking the system.

\subsection{The Process of Change Impact Analysis}

When a change occurs in a single repository, the ISAR process maps the changed components to a delta, which scopes the direct impact on the intermediate representation. However, the intermediate representation connects components within and beyond a single microservice, and thus, a potential indirect impact on the overall system (other microservices) can be determined without the need to engage other microservice repositories in the process.

This has multiple consequences; a calculated direct impact of a delta can encapsulate the architectural impact implied by the intermediate representation since it can also be used to derive architectural descriptions \cite{cerny24cluster}. The directly impacted components in the intermediate representation are interconnected with other components, and these can be traversed through the component call graph alongside microservice dependencies to determine the indirect impact. This provides a mechanism to determine the potential span of the change to other microservices and their specific components. 

\begin{figure}[t!]
    \centering
     \vspace{-0.7em}
    \includegraphics[width=23.8em]{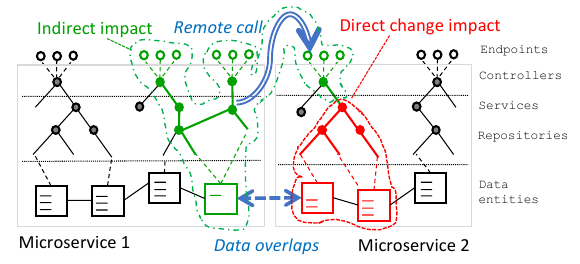}
    \vspace{-0.7em}
    \caption{Illustration of change impact analysis, red bounded section marks the changes in service, two repositories, and two data entities, and the green bounded section shows the potential area for indirect impact.}
    \label{fig:ciasample}
    \vspace{-1.5em}
\end{figure}

Figs. \ref{fig:isar} and \ref{fig:ciasample} illustrate a delta impact with a direct change impact in components highlighted within red boundaries (for Fig. \ref{fig:ciasample} it is a service, two repositories, and two data entities). However, there is also a potential indirect impact that needs to be reviewed. Modified components might be used by other components (i.e., controllers using a modified service) or have dependencies beyond the actual microservice and cascade through the internal components. This is illustrated within green boundaries. 
 


\begin{wrapfigure}{r}{0.285\textwidth}
   \vspace{-0.3em}
    \centering
    \includegraphics[width=11em]{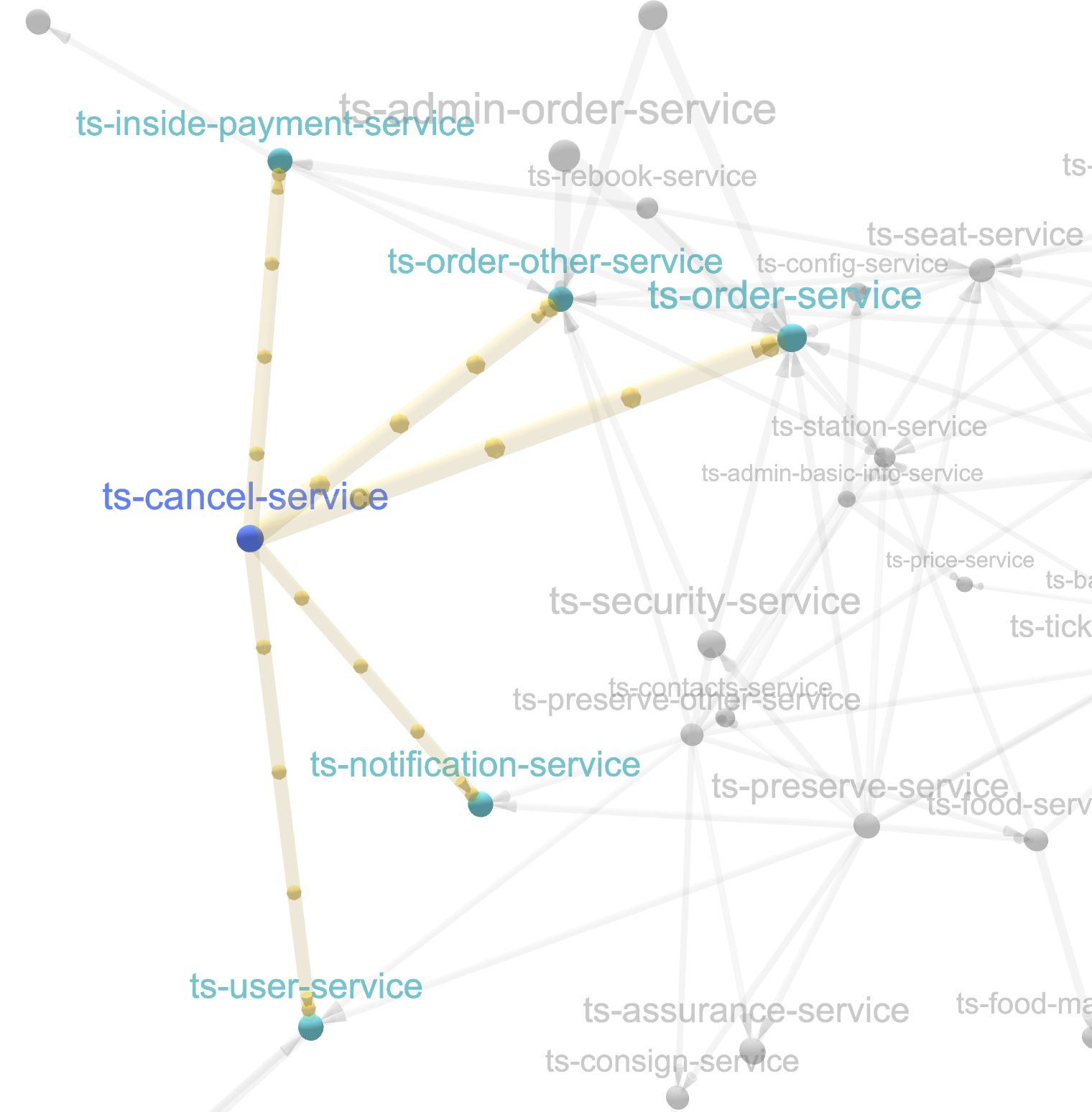}
        \vspace{-0.7em}
    \caption{Illustrated integration of change impact analysis to interactive models \cite{10.1007/978-3-031-61816-1_15,10.1007/978-3-031-66326-0_19}}
    \label{fig:changeImpact}
    \vspace{-.4em}
\end{wrapfigure}

A change with a potential impact on another microservice might signal the need to engage the particular microservice's team in a change review. Change impact is available in simple text reports that might not be too descriptive to developers about changed microservices. With the visual integration into interactive models \cite{10.1007/978-3-031-61816-1_15,10.1007/978-3-031-66326-0_19} we can highlight changes span to other microservices, see Fig. \ref{fig:changeImpact}. However, these interactive views need to become hierarchical to enable the detailed observation of components and how they are connected to illustrate potential change propagation pathways.






\subsection{Narrowing Detection of Breaking Changes and Conflicts}


Changes possibly cascading across microservices need to be alerted to involve relevant development teams. This is especially important because of the impact that change could have through indirect effects. However, a plain illustration of potential propagation pathways of an indirect impact might mislead reviewers without an actual effect. Providing the ability to design more specific rules tailored to desired situations that matter in a review could be a powerful instrument to alert practitioners about potential conflicts that could happen as an effect of a change. These rules should be descriptive to identify the changed and also impacted components with the granularity of each change and alert potential issues to review.

The ISAR infrastru-cture enables us to approach the detection of potentially breaking changes and conflicts. The intermediate representation is a hierarchical structure that can be traversed as a graph and can integrate with delta content (JSON formats). The traversable graph format enables the identification of various graph properties and, thus, a setting of custom rule definitions. These rules (to narrow potential breaking changes and conflicts) can be arbitrary, and the infrastructure enables checking such rules throughout the system's evolution.


ISAR, as illustrated in Fig.\ref{fig:ToolFlow}, incrementally adds deltas to the system's intermediate representation baseline; this is where to hook to the process with potential conflict checks performed upon deriving increments (3). The inputs to deriving an increment are a delta and a system baseline. 
Thus, the potential conflict rules might use information regarding the change in the system version, while others might be identified by a complete system version. This difference is a foundational idea of our approach, as the encapsulated version change is more efficient for decentralized environments. 
The change-based approach saves time and resources, avoiding scanning the entire system; moreover, it is possible to tie additional rules broken to specific code changes made directly.


An example rule that would work with the delta is when a service method gets modified, impacting its return object, which poses a risk to the service method's parent microservice and all consumer services. With the context of a delta, it is possible to identify modifications that generate risk to the dependencies established between microservices. When a potential conflict is found throughout a microservice change, it can be reported with relevant contextual information, such as the problem component or other affected components, to facilitate the review process.

  

To illustrate more examples, 
consider the addition of a rest call to a microservice, which targets an endpoint that is not present in the system. Another example is the breaking of a link between \texttt{microservice A} and \texttt{microservice B} by removing an endpoint in \texttt{microservice A}, which removes the link between the two services performed in some business logic of the service layer. The complexity of the rules could vary from case to case.

The rule concept to detect change introducing potential conflicts considers various aspects. Listing \ref{lst:RuleOutline} gives a rule outline that indicates whether the analysis takes place over the delta or the system context; next, it gives specific component types involved along with the focus on the change type. Consequently, the monitored impact is specified in terms of component types and impact types. Using the concept of delta or the system and marked component nodes in the connected graph makes it flexible to set up arbitrary graph traversal rules to detect potential conflicts across changes.

\lstdefinestyle{mystyle}{
    backgroundcolor=\color{backcolour},   
    commentstyle=\color{codegreen},
    keywordstyle=\color{magenta},
    morekeywords={Name,Change,Condition,Node,Scope, Change, Impact, Confidence,WC, Hub, AL, AR, UE,IC,FC, Floating~Call,AnalysisLevels,ChangedComponents,ChangeType,MonitoredImpact,ComponentType,ImpactType}, 
    numberstyle=\tiny\color{codegray},
    stringstyle=\color{codepurple},
    basicstyle=\ttfamily\scriptsize,
    breakatwhitespace=false,         
    breaklines=true,                 
    captionpos=b,                    
    keepspaces=true,                 
    numbers=left,                    
    numbersep=5pt,                  
    showspaces=false,                
    showstringspaces=false,
    showtabs=false,                  
    tabsize=2,
    belowcaptionskip=1em,
    belowskip=1em,
}

\lstset{style=mystyle}

\begin{lstlisting}[caption=Potential Change Conflict Rule Outline,label=lst:RuleOutline]
{ 
  AnalysisLevels: [Delta | System], 
  ChangedComponents: [
    {ComponentType: [Endpoint | Call | Controller 
                          | Service | Repository],
     ChangeType: [Delete | Update | Add]}
  ], 
  MonitoredImpact: {
     ComponentType: [Endpoint | Call | Controller 
                          | Service | Repository],
     ImpactType: Unused | Inconsistent | Unmatched}
}
\end{lstlisting}

With this process in place for the evolution of the entire system, decentralized team changes become more controlled with regard to the impact on the overall system, which could otherwise go unseen in local repository review. It would help developers and facilitate the review process of determining the potential scope of local changes, as well as inviting relevant microservice team representatives for the change review to ensure a smooth new version rollout. 

\section{Case Study}

    


To assess the proposed infrastructure and change impact analysis approach, we implement a proof-of-concept (PoC) tool\footnote{Find our open-source tool at $<$URL BLIND REVIEW Anonymized$>$}. To ensure the reliability of the ISAR process, we based our tool on the existing tool Prophet \cite{cerny24cluster} for the SAR since it has been manually validated \cite{hutcheson2024software} on the train-ticket benchmark \cite{Zhou2018}. However, we extended the capabilities to recognize additional concepts across Java frameworks and implemented the ISAR with deltas. In addition, we also implemented change impact analysis based on the concepts from Section IV.

\subsection{Considered Potential Change Conflict Situations}

To assess our potential change conflicts rule approach, we considered and implemented sample rules, summarized in Listing \ref{lst:RuleDef}. When conditions of a particular rule are found, we flag it for the increment review process as a potential conflict alongside the change impact context within relevant components. 
The considered rules are detailed below.


\lstset{moredelim=[is][\bfseries]{[*}{*]}}
\begin{lstlisting}[caption= Potential Change Conflict Detection Rule Outline,label=lst:RuleDef,]
[*Invalid Call (IC):*] A rest call targets a non-existent endpoint
[*Uncalled Endpoint (UEM):*] An endpoint present in the system is not called from another middleware microservice
[*Service Method Modified (SMM):*] A service method is modified and possibly returns inconsistent results
[*Repository Method Modfied (RMM):*] A repository method is modified and possibly returns inconsistent results
\end{lstlisting}

An \textit{Invalid Call (IC)} targets an endpoint that is not present in the overall system. This can be detected by scanning added or modified REST calls against the overall system for a match; if no match is found, it is flagged. Any business logic, depending on the results of this REST call, is at risk of providing faulty or no results to end users or the application.

\begin{lstlisting}[caption=Invalid Call (IC),label=lst:IC]
{ 
  AnalysisLevels: ["System"], 
  ChangedComponents: [
   {ComponentType: ["Endpoint", "Call"],
    ChangeType: ["All"]}], 
  MonitoredImpact: {
    ComponentType: "Call",
    ImpactType: "Unmatched"}
}
\end{lstlisting}


An \textit{Uncalled Endpoint From Middleware (UEM)} is when an endpoint advertised in a \texttt{microservice A} is not utilized by any other middleware microservice in the system. We can detect this by iterating over all collected endpoints, and if no rest call matches that endpoint, we make a detection. Since this adds additional complexity to the codebase and demands additional overhead and maintenance, it is flagged. This rule might give a ghost prediction if the endpoint is used by 3rd party system or called from a user interface. 

\begin{lstlisting}[caption=Uncalled Endpoint (UEM),label=lst:UE]
{ 
  AnalysisLevels: ["System"], 
  ChangedComponents: [
   {ComponentType: ["Endpoint", "Call"],
    ChangeType: ["All"]}], 
  MonitoredImpact: {
    ComponentType: "Endpoint",
    ImpactType: "Unused"}
}
\end{lstlisting}

A \textit{Service Method Modification (SMM)} is when there is a modification to the data returned from a service layer method, and it returns inconsistent results from the previous version. This rule is detected by scanning for method calls on an object of the same type as the defined return type for the service method. This method is at risk of providing inconsistent results to the controller layer, affecting endpoint consumers.

\begin{lstlisting}[caption=Service Method Modified (SMM),label=lst:SMM]
{ 
  AnalysisLevels: ["Delta"], 
  ChangedComponents: [
    {ComponentType: ["Service"], 
     ChangeType: ["Modify"]}], 
  MonitoredImpact: {
    ComponentType: "Service", 
    ImpactType: "Inconsistent"}
}
\end{lstlisting}

A \textit{Repository Method Modified (RMM)} comes with a modification to the definition of a repository layer method. This rule is detected by scanning for \texttt{[ADD/MODIFY/DELETE]} annotations on repository layer methods. This method might bring a risk of providing inconsistent results to the service layer and affect calls and endpoints. 

\begin{lstlisting}[caption=Repository Method Modified (RMM),label=lst:RMM]
{ 
  AnalysisLevels: ["Delta"], 
  ChangedComponents: [
   {ComponentType: ["Repository"],
    ChangeType: ["Modify"]}], 
  MonitoredImpact: {
    ComponentType: "Repository",
    ImpactType: "Inconsistent"}
}
\end{lstlisting}

\subsection{Validation Project Dataset}

The project benchmarks we used for the evaluation were systematically selected from the microservice dataset provided by
Amoroso et al. \cite{MSR2024}. We filtered projects that are based on Java Spring Boot Framework and have more than one dependency between services. For instance, many project were not interconnected through remote calls. The resulting validation project set counts 8 benchmarks provided in Table~\ref{tab:microProjects}.

\begin{table}[h!]
\centering
\caption{Validation microservice projects.}
\label{tab:microProjects}
\vspace{-.5em}
\begin{tabular}{@{}llr} 
 \hline
\textbf{ID}&  \textbf{Repository (at https://github.com/..) }& \textbf{\#Commits}  \\ 
 \hline
P1& ../FudanSELab/train-ticket & 323  \\

P2& ../apssouza22/java-microservice & 140 \\
 
P3& ../shabbirdwd53/Springboot-Microservice & 7 \\ 
 
P4& ../mdeket/spring-cloud-movie-recommendation  & 6 \\

 P5&../piomin/sample-spring-microservices-new & 181  \\

P6& ../koushikkothagal/spring-boot-microservices-workshop & 12 \\

P7& ../anilallewar/microservices-basics-spring-boot & 46  \\

 P8&../AthirsonSilva/spring-microservices & 26  \\

 \hline
\end{tabular}

\end{table}

\subsection{Evaluation}

Our tool's ISAR process was executed on the validation project dataset code repository history. We verified that increments from combining deltas to the intermediate representation baseline produce the same next intermediate representation version as if generated by a complete SAR process. 

The change impact evaluation using rules to target situations with breakage potential considered (1) injected change anomalies to their repository and (2) manual validation of changes identified by the rules in the natural evolution of the projects.

\subsubsection{Anomaly injection} 

Our initial change impact evaluation was simulated with injected anomalies. In particular, we took the train-ticket benchmark (P1) \cite{Zhou2018} and fabricated commits with breaking changes and conflicts. We manually injected at minimum one anomaly for each rule to our test project.
These anomalies covered a variety of cases, including modifications to rest calls and endpoints and changes to static implementation files. We achieved this through unit testing to ensure each rule in isolation to uphold accuracy. This approach allowed for the manual validation of our rules against our approach. Next, we validated that our tool detects the potential conflict changes through the implemented rule detection. The results concluded we successfully implemented our approach for each conflict rule and served as a test suite.





\subsubsection{Non-altered history analysis}

Next, we resorted to manual validation using the non-altered evolution history of the validation project dataset. Given the vast scope of dataset history, we resorted to a reduced scope. In particular, we focused on the train-ticket benchmark (P1), where we manually identified and labeled 25 changes with breaking potential or potential of causing consistency error. Consequently, we checked that our tool identified the changes. Our tool identified all 25 instances. 

\begin{figure*}[t!]
\begin{minipage}{.38\textwidth}

\tikzset{>=stealth}
\begin{tikzpicture}[scale=0.70]
\begin{axis}[legend entries={Invalid Call, Uncalled Endpoint from Middleware, Service Method Modified,  Repository Method Modified},
legend style={font=\scriptsize},
 height=6cm,
width=10cm,
xlabel=Commit History in P1 (train-ticket),
legend style={at={(.99,0.65)}},
    ylabel=Rule Violation Per Version]
\addplot[solid, red] table [x=Index, y=AR1, col sep=comma] {data/train-ticket.csv};
\addplot[solid, orange] table [x=Index, y=AR2, col sep=comma] {data/train-ticket.csv};
\addplot[solid, green] table [x=Index, y=AR3, col sep=comma] {data/train-ticket.csv};
\addplot[solid, blue] table [x=Index, y=AR4, col sep=comma] {data/train-ticket.csv};
\end{axis}
\end{tikzpicture}

\end{minipage}
\begin{minipage}{.29\textwidth}

\tikzset{>=stealth}
\begin{tikzpicture}[scale=0.70]
\begin{axis}[legend entries={Invalid Call, Uncalled Endpoint from Middleware, Service Method Modified, Repository Method Modified},
legend style={font=\scriptsize},
 height=6cm,
width=8cm,
legend style={at={(.99,0.65)}},
xlabel=Commit History in P2 (java-microservice),
    ylabel=Rule Violation Per Version]
\addplot[solid, red] table [x=Index, y=AR1, col sep=comma] {data/java-microservice.csv};
\addplot[solid, orange] table [x=Index, y=AR2, col sep=comma] {data/java-microservice.csv};
\addplot[solid, green] table [x=Index, y=AR3, col sep=comma] {data/java-microservice.csv};
\addplot[solid, blue] table [x=Index, y=AR4, col sep=comma] {data/java-microservice.csv};
\end{axis}
\end{tikzpicture}
\end{minipage}
\begin{minipage}{.31\textwidth}

\tikzset{>=stealth}
\begin{tikzpicture}[scale=0.70]
\begin{axis}[legend entries={Invalid Call, Uncalled Endpoint from Middleware, Service Method Modified, Repository Method Modified},
legend style={font=\scriptsize},
 height=6cm,
width=8cm,
legend style={at={(.99,0.65)}},
xlabel=Commit History in P5 (sample-spring-microservices-new),
    ylabel=Rule Violation Per Version]
\addplot[solid, red] table [x=Index, y=AR1, col sep=comma] {data/sample-spring-microservices-new.csv};
\addplot[solid, orange] table [x=Index, y=AR2, col sep=comma] {data/sample-spring-microservices-new.csv};
\addplot[solid, green] table [x=Index, y=AR3, col sep=comma] {data/sample-spring-microservices-new.csv};
\addplot[solid, blue] table [x=Index, y=AR4, col sep=comma] {data/sample-spring-microservices-new.csv};
\end{axis}
\end{tikzpicture}
\end{minipage}
    \vspace{-0.3em}
  \caption{P1/P2/P5 Rule Violations over History}
     \label{fig:TrainticketHistory}
     \vspace{-1.3em}
     
\end{figure*}
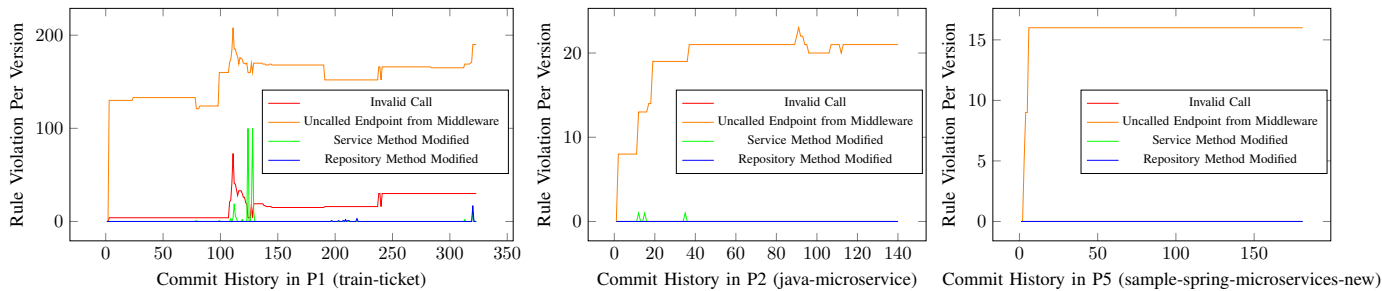

The next assessment considered the precision of our identification. For each benchmark, we mined the complete commit history and natural evolution flow to analyze conflict rules over time. We recovered a dataset of potentially conflicting changes, which we share through a Zenodo package\footnote{Zenodo package at \url{https://zenodo.org/records/13922262}
contains the full dataset of artifacts generated by our tool for assessed microservice benchmarks; for ISAR, it contains the list of intermediate representations with deltas per project; it also shares identified violations and our validation.   
}. 
Table \ref{tab:microProjectsVio} outlines the total number of unique violations of each rule found in the history across the considered project.
Consequently, two developers assessed the entire dataset of potentially conflicting changes with the best effort. While a broad effort was invested in the assessment, this does not imply we detected potential rule violations of given rule types.


\begin{table}[t!]
\centering
\caption{total number of unique violations of each rule defined in Listings \ref{lst:RuleDef} found in the history across the considered project indexed in Table \ref{tab:microProjects} (see Zenodo package$^2$ for a detail)}
\vspace{-.5em}
\label{tab:microProjectsVio}
\begin{tabular}{lrrrrr} 
 \hline
\textbf{Project} &  \textbf{IC}& \textbf{UEM}& \textbf{SMM}& \textbf{RMM} & \textbf{\#Commits}  \\ 
 \hline
P1& 137& 302& 273& 27 & 323\\

P2& 0& 21& 3& 0 & 140 \\
 
P3& 0& 5& 0& 0 & 7 \\
 
P4& 1& 9& 0& 0  & 6\\

 P5& 0& 16& 0& 0 & 181 \\

P6& 1& 3& 0& 0  & 12\\

P7& 1& 7& 0& 0  & 46\\

 P8& 0& 8& 1& 0  & 26 \\

 \hline
\end{tabular}
\vspace{-1em}
\end{table}


\subsection{Rule Violation Dataset Evolution Analysis and Discussion}

Most of the rule violations came from the uncalled endpoint of middleware (UEM). Uncalled endpoints may not be so surprising because it is common for developers to generate endpoints before the creation of calls. Unlikely, though, is the generation of rest calls that do not target any endpoint present in the system. Since we scan from a system perspective, we identify bad calls that result from the deprecation of API endpoints or modification of these endpoints. At the same time, when we dive more deeply into the train-ticket benchmark (P1) and consider existing evaluation works \cite{electronics13101913}, there were findings identifying 262 endpoints in the project at one of its versions. When running best-effort coverage end-to-end tests on the user interface, 119 endpoints were identified. That leaves many potential endpoints unused in the system. Moreover, the same study suggests that 39 endpoints are isAlive methods to ensure the system is running. While our tool did not analyze the user interface, there can be ghost situations where 3rd party system calls the endpoint, or the user interface calls it. While not a limit of the approach, the lack of user interface code analysis is a limit of our tool.

The second most common rule violation came from service method modifications (SSM), which are changes to the return object of a service method. This is generally to be expected; the manipulation of data is done in the business layer, so naturally, it holds all the logic of the API.

These manipulations are critical because when return data from the business layer is modified, it ultimately changes the behavior of the API and dependent endpoints. This connection between the controller layer and the service layer is what drives this architectural rule. It is a poor assumption to say service \texttt{method A} and \texttt{method B} return identical results if \texttt{method B} suddenly has two additional method calls on the return object. 

Other common rule violation include invalid calls (IC). Since we scan from a system perspective, we identify bad calls that result from the deprecation of API endpoints or modification of these endpoints. When cross-analyzing with code changes, this is exactly what is happening in the code; modifications to annotations and endpoint paths cause these calls to target invalid or non-existent endpoints.

The repository method modification (RMM) indicated a rule violation solely in the train-ticket benchmark. It is consistent with changes to repository components, specifically when annotations are added or removed with method definitions. Most often, this happened to be the removal of custom Query annotations in exchange for default JPA or CRUD implementations. This may cause different behavior or results from what is previously expected in the service layer. It's important to track these changes as they could change the execution flow of the service layer or the data returned to the controller layer.

Figure \ref{fig:TrainticketHistory} breaks down the history of rule violations over time, showing three of the 8 projects that have the longest commit history (P1, P2, and P5). As mentioned previously, some rule violations are identified from complete increments of the system's intermediate representation, while others are based on changes in the system (through a baseline and a delta). Therefore, we find two varying lines determined by delta tracking or system tracking. Narrow peaks and valleys are indicative of delta tracking, where the number of rule violations will shoot up over some delta change and return to a baseline of zero. Otherwise, the remaining lines portray a more steady moving average of rule violations found in the system as modifications are made over time. These modifications alter the overall standings of the system yet can not be detected from the delta alone, so the entire system is scanned at each merge to identify its current standings.

Interpreting the data, shifts in degradation are assumed by high activity in the graph. For example, looking at the period between 100 and 150 commits for train-ticket (P1), we can identify a lot of activity as the number of rule violations dramatically increases and decreases. When manually identifying the code between these commits, we see large refactoring and additions made that introduce new endpoints and other calls that do not immediately match up. Additionally, we find a large number of changes to business logic causing the jumps in service method modifications (SMM).

\subsection{Implications}
Our PoC demonstrates the capability to execute ISAR on sample projects and illustrates how the resulting intermediate representation and deltas can be utilized for change impact analysis involving decentralized microservices. In the context of existing practitioner surveys \cite{lercher2024microservice,bavskarada2020architecting,bogner2021} on related microservice-based system challenges, our approach should be seen as transformational with broad potential in the community. While others \cite{10.1007/978-3-031-70797-1_10} might focus on a shallow perspective considering only API evolutions, our approach goes down to the component level.

Our assessment indicates the suitability of the approach to change impact analysis in multiple microservice projects, with the granularity of components. 
It demonstrates the ability to narrow situations with the potential for breaking changes or consistency errors through customizable rules that use the ISAR intermediate representation and deltas. Yet, these rule violations do not necessarily indicate a breakage but aim to flag attention in the review process.
The approach can be applied to development pipelines to bring practical implications. The variability of the assessed project reassures broader applicability.  

Identified rule violations provide instant feedback to developers but should be considered with reservations because the approach is based on static analysis, which leads to approximation. 
In the instance of ISAR, the approach has the potential to measure quantitative shifts in architecture since the intermediate representation used in our work can be used to derive architectural views and descriptions. This leads to an opportunity to assess the architectural impact of a particular commit of the pull request. Given our approach is extensible, deltas could collect commit metadata and assess developers based on their architectural impact. Given that the intermediate representation has the granularity of components, the evolution history could collect specific developers that contribute the most to the component life-cycle, and this directly indicates who should be involved in the change review for changes that are promoted across microservices. 

The entire approach can be connected to the established practice of smell detection and technical debt; however, our work intended to provide an alternative direction that we believe has finer granularity and greater potential to guide developers in the observability of changes and change impact estimations in the system composted from many microservices that they lack proper understanding of, which is a recognized challenge \cite{lercher2024microservice,bavskarada2020architecting,bogner2021}.
%
%
Developers are typically unaware of the overall system dependencies, but with our approach, their individual change to a particular microservice becomes better traceable and gains an estimated impact on the overall system, giving immediate feedback. Something like this is currently missing in the market for microservices to guide throughout the system evolution management.

\subsection{Threats to validity}

\noindent
We\,address the threats to validity outlined by Wohlin\,et\,al.~\cite{10.5555/2349018}.

\subsubsection{Construct Threats}
Our approach assumes access to the system’s source code and changes. To mitigate this, we rely on version control systems, which track project history and modifications, ensuring our methodology aligns with real-world practices. The study did not consider user interface.



\subsubsection{Internal Threats}
A potential threat is the data used for validation. To minimize errors, we split the authors into two groups: one for information extraction and the other for independent validation, such as labeling rule violations in the train-ticket benchmark and verifying identified violations in the validation dataset with our PoC tool.



\subsubsection{External Threats}

A key threat is the generalization of our approach to different microservice-based systems. While the prototype was implemented for Java Spring projects, we mitigated this by evaluating it on eight distinct projects. Additionally, the methodology itself is designed to be language-agnostic and adaptable to various project structures. The rules used are flexible and not limited to the examples shown, which were chosen to illustrate a variety of potential situations. However, further assessment of real-world production systems is needed for broader validation. Additionally, empirical studies with practitioners are necessary to further validate the method's effectiveness and we share generated artifacts$^2$.

\subsubsection{Conclusion Threats}
A potential threat to conclusion validity is the absence of a benchmark dataset for comparison, given the novelty of the proposed method. To mitigate bias, we validated results with independent local teams. Additionally, performance concerns are addressed by following the ISAR and calculating delta changes, merging them with the previous IR state to optimize efficiency, avoiding the heavier approach of comparing system snapshots at different stages.




\section{Conclusion}

This paper introduced a new infrastructure that aims to advance change impact analysis in microservices. To foster advancements in the maintainability of cloud-native systems, we presented a novel approach using ISAR that is based on intermediate representation and deltas.
Our advancement has the potential to mitigate ripple effects and assist practitioners with local changes with unforeseeable broader implications on the system.
The most likely adoption of the approach is for development pipelines, yet our PoC tool is intended to demonstrate the capability of change impact analysis rather than to offer a minimum viable product ready for customers. 

To promote community advancements, we shared our tool and a complete data set of artifacts generated in our study, including intermediate representations and deltas of 8 projects throughout their evolution and a dataset of identified potential change conflicts. These artifacts could be used in consequent empirical studies or validations.


The presented perspective aims at the versatility of the approach, which uses the traditional static code analysis method and utilizes microservice dependencies, facilitating its future extensions. While one should recognize the limits of approximation inherited from the static analysis approach, one should also see the broader potential for adoption and extension. New tools could emerge to visualize the impact of change and inform reviewers about estimated consequences. 


In the context of the recognized challenges with microservice system evolvability \cite{bogner2021} and breaking changes \cite{lercher2024microservice}, the current state-of-the-art focus is on API evolution. Our approach, showcases advancement through analysis beyond the API, targeting the component level, which provides great context when indicating change impact. It intends to better manage quality assurance throughout system evolution. It has the potential to coordinate decentralized teams to mitigate ripple effects. 

Our future work is multi-fold. We plan for validation of the actual impact with developers. Also, we plan to extend the visualization capabilities to add an interactive component view to highlight change impact. We plan to aggregate developer karma for each component across the delta history to identify the main contributors to a given component that is involved in change propagation. Finally, we aim to research change impact summarization to mitigate communication overhead across decentralized teams working on different microservices. In particular, the ISAR infrastructure could be used with Large Language Models (LLM) to summarize the actual change details, similar to Quevedo et al. \cite{quevedo2024evaluating}, who tried to summarize endpoints and other system components and methods. Finally, there is a promising direction to deal with heterogeneity, which involves language-agnostic intermediate representation, which could be reconstructed across various platforms \cite{schiewe2022advancing}.

\section*{\small Replication Package and Artifacts}
{\scriptsize
Full
dataset of artifacts generated by our tool; lists of intermediate representations and deltas
per benchmarks; it also shares identified violations and our validation. \url{https://zenodo.org/records/13922262}.
}

\section*{\small Acknowledgement}
{\scriptsize
This material is based upon work supported by the National Science Foundation under Grant No. 2409933. Any opinions, findings, and conclusions or recommendations expressed in this material are those of the author(s) and do not necessarily reflect the views of the National Science Foundation.
}

\bibliographystyle{IEEEtran}      
\bibliography{bib}   

\end{document}